\documentclass[12pt,a4paper]{article}
\usepackage[latin1]{inputenc}
\usepackage{amsmath}
\usepackage{amsthm}
\usepackage{amsfonts}
\usepackage{amssymb}
\usepackage{hyperref}
\usepackage{cleveref}
\usepackage{graphicx}
\usepackage{nicefrac}
\usepackage{fullpage}
\usepackage{bm}
\usepackage{xcolor}
\usepackage{tikz}
\usepackage{thmtools}
\usepackage{thm-restate}

\usepackage{mathtools}
\usepackage{algpseudocode}

\newcommand\blfootnote[1]{%
  \begingroup
  \renewcommand\thefootnote{}\footnote{#1}%
  \addtocounter{footnote}{-1}%
  \endgroup
}

\newtheorem{thm}{Theorem}[section]

\newtheorem{lemma}[thm]{Lemma}
\newtheorem{remark}[thm]{Remark}
\newtheorem{cor}[thm]{Corollary}

\newtheorem{claim}[thm]{Claim}

\newtheorem{obser}[thm]{Observation}
\newtheorem{example}[thm]{Example}

\newcommand{\Fq}{\mathbb{F}_q}

\newcommand{\cC}{\mathcal{C}}

\makeatletter

\newcommand{\Rmnum}[1]{\expandafter\@slowromancap\romannumeral #1@}
\makeatother

\author{
     Itzhak Tamo \thanks{
      Itzhak Tamo is with the Department of Electrical Engineering--Systems, Tel Aviv University, Tel Aviv, Israel. e-mail:	 tamo@tauex.tau.ac.il.} }

\begin{document}
\title{Tighter List-Size Bounds for List-Decoding and Recovery of Folded Reed-Solomon and Multiplicity Codes}
\maketitle
\begin{abstract}
Folded Reed-Solomon (FRS) and univariate multiplicity codes are prominent polynomial codes over finite fields, renowned for achieving list decoding capacity. These codes have found a wide range of applications beyond the traditional scope of coding theory. In this paper, we introduce improved bounds on the list size for list decoding of these codes, achieved through a more streamlined proof method. Additionally, we refine an existing randomized algorithm to output the codewords on the list, enhancing its success probability and reducing its running time. Lastly, we establish list-size bounds for a fixed decoding parameter. Notably, our results demonstrate that FRS codes asymptotically attain the generalized Singleton bound for a list of size $2$ over a relatively small alphabet, marking the first explicit instance of a code with this property.

\end{abstract}

\section{Introduction}
\blfootnote{This work was supported by the European Research Council (ERC grant number 852953).}
An error-correcting code \( C \subseteq \Sigma^n \) is a subset of vectors, known as codewords, each of length \( n \) over a specified alphabet \( \Sigma \). The primary goal in constructing such a code \( C \) is to enable the retrieval of an original codeword \( c \in C \) from its corrupted version, while also maximizing the size of \( C \). In the standard problem of unique decoding, the objective is to accurately reconstruct \( c \) from any corrupted version \( \tilde{c} \in \Sigma^n \) that differs from \( c \) in less than \( \delta n/2 \) positions. This requires the code's relative distance, defined as the minimum proportion of differing positions between any two distinct codewords, to be at least \( \delta \).

However, in both coding theory \cite{elias2006errorcorrecting,blinovskii1986bounds,ahlswede1973channelcapacities}) and complexity theory \cite{sudan2000listdecoding,vadhan2012pseudorandomness}, the challenge of decoding from a higher level of corruption, specifically beyond the \( \delta n/2 \) threshold, has gained prominence. Here, \( \delta \) represents the code's relative distance. In such cases, known as list decoding 
(first considered  by Elias \cite{peter1957listdecoding} and Wozencraft \cite{wozencraft1958listdecoding}), 
unique decoding becomes infeasible, and the focus shifts to identifying a short list of codewords of \(  C \) that includes the original codeword \( c \).

Central to the study of list decoding are algebraic codes, which serve as the primary examples of codes that can be list-decoded efficiently. A particularly notable example is the family of Reed-Solomon (RS) codes \cite{reed1960polynomial}, which are  formed by the evaluations of polynomials of low degree. During the 1990s, these codes were demonstrated to have efficient list-decoding algorithms. Initially, Sudan's work \cite{sudan1997decoding} showed that RS codes are indeed list decodable beyond half of their minimum distance. This result was later improved by Guruswami and Sudan \cite{Guru-sudan-algo}, where they developed an  algorithm, now known as the Guruswami-Sudan algorithm, capable of list-decoding up to the Johnson radius. These significant advancements have spurred the development of the field of algorithmic list decoding, and indeed more recently variations of RS codes were shown to have enhanced list  decoding capabilities. This paper focuses on two such variations: Folded Reed-Solomon (FRS) codes and multiplicity codes. 

Folded Reed-Solomon (FRS) codes, constructed by Guruswami and Rudra in \cite{GR08}, represent a simple variation of RS codes. Consider an  RS codeword as $(c_0, c_1, \ldots, c_{n-1}) \in \Sigma^n$. Its folded version with  a folding parameter $s$, is structured as follows:
\[
\begin{bmatrix}
    c_0 & c_1 & \ldots & c_{s-1} \\
    c_s & c_{s+1} & \ldots & c_{2s-1} \\
    \vdots & \vdots & \ddots & \vdots \\
    c_{n-s} & c_{n-s+1} & \ldots & c_{n-1}
\end{bmatrix} \in (\Sigma^s)^{\frac{n}{s}}.
\]
The significance of FRS codes lies in their enhanced capabilities for list decoding beyond the Johnson radius. FRS codes, in particular, allow for a significantly higher error (corruption) tolerance for a given code rate, achieving an asymptotically optimal threshold, termed as the list-decoding capacity.

Multiplicity codes, introduced by Rosenbloom and Tsfasman \cite{rosenbloom1997codes}, extend polynomial codes by incorporating evaluations of polynomials and their derivatives. In a multiplicity code, a symbol is represented as \((f(x), f^{(1)}(x), \ldots, f^{(s-1)}(x)) \in \mathbb{F}_q^s\), where \(f^{(i)} \in \mathbb{F}[X]\) is the \(i\)th derivative of a low-degree polynomial \(f\) and \(x \in \mathbb{F}_q\). Similar to FRS codes, multiplicity codes also achieve list decoding capacity, as demonstrated in \cite{kopparty2015list,guruswami-wang}.

Historically, the output list size of these codes was known to be polynomial in the code's length. However, recent advancements by Kopparty et al. \cite{kopparty-improved-list-size} have shown that these codes possess much smaller list sizes than previously known. Specifically, they demonstrated that these codes achieve list decoding capacity with constant list sizes, independent of the block length. This contrasts with earlier works where constant list sizes were achieved only by relying on pre-encoding with subspace evasive sets \cite{guruswami-wang,dvir2012subspace}

In this paper, we present an improvement over the results of Kopparty et al. \cite{kopparty-improved-list-size} by establishing a tighter bound on the list size for both Folded Reed-Solomon (FRS) codes and multiplicity codes. Our approach, while bearing similarities to that of \cite{kopparty-improved-list-size}, is distinguished by its simplicity. Despite these similarities, our approach enables us to refine and improve upon their established bounds. Our contribution is summarized as follows:

\begin{itemize}
    \item We present improved bounds on the list size of list decoding FRS and multiplicity codes, employing a more streamlined proof.
    Moreover, for FRS codes, we provide a more general result regarding list recovery, and for this case, we also establish a tighter bound on the list size. Lastly,  we slightly modify a randomized algorithm of  \cite{kopparty-improved-list-size} that outputs the list for the list recovery problem with high probability. The modification leads to  a faster running time and higher success probability of the algorithm. 
    \item We establish bounds on the list size in list decoding FRS codes for a fixed decoding parameter \(m\) (as defined in Theorem \ref{guruswami-thm}). Specifically, for \(m=2\), we demonstrate that the bound is tight in the sense that the decoding radius asymptotically achieves the generalized Singleton bound \cite{Shangguan2020combinatorial} for a list of size \(2\), over a relatively small alphabet. This represents the first known instance of list decodable codes achieving this property.
\end{itemize}

\subsection{ Overview of Main Results:}

\vspace{0.2cm}
\noindent{\bf Enhanced List Decoding of FRS and multiplicity Codes.}
We begin with  some needed notation. The rate of a code $C \subset \Sigma^n$ is defined as $R := \frac{1}{n} \log_{|\Sigma|} |C|$. The minimum fraction of differing coordinates between any two codewords in $C$ is called  the relative distance of $C$ and is denoted by $\delta$. A code $C$ is called $(\rho, L)$-list decodable if for any received word $w \in \Sigma^n$, there exist no more than $L$ codewords $c \in C$ such that $c_i = w_i$ for all but a $\rho$ fraction of coordinates $i$. 
Similarly, $C$ is called $(\rho,\ell, L)$-list recoverable
if for any  $n$ sets $S_i\in \Sigma$ of size $\ell$ there are at most $L$ codewords $c \in C$ such that $c_i \in S_i$ for all but a $\rho$ fraction of the $i$'s. For both definitions  $\rho$ is called the decoding radius.

The well-known Guruswami-Sudan list decoding algorithm \cite{Guru-sudan-algo} is capable of efficiently list decoding RS codes of rate $R$ up to a radius $\rho= 1 - \sqrt{R}$, with polynomial list sizes, thus achieving  the Johnson bound.
Codes that can surpass this decoding radius while maintaining polynomial or even constant list sizes for large alphabets were known to exist. The theoretical limit for list decoding any code of rate $R$, known as the list decoding capacity, is $\rho=1-R$, achieved  by  random codes. Specifically, a random code with rate $R$ and alphabet size $\exp(1/\varepsilon)$ is $(1 - R - \varepsilon, O(1/\varepsilon))$-list decodable with high probability. However, explicitly constructing codes that could efficiently reach list decoding capacity was a major open question.

In a breakthrough result by  Guruswami and Rudra \cite{GR08}, building upon Parvaresh and Vardy's work \cite{parvaresh2005correcting}, it was shown that the folding operation of RS codes can improve their list decoding capability up to capacity with  polynomial list sizes. Formally, \cite{GR08} proved that an FRS code with rate $R$ and folding parameter $s \approx 1/\varepsilon^2$ is $(1 - R - \varepsilon, n^{O(1/\varepsilon)})$-list decodable, which means polynomial in $n$ list size. Few years later,  multiplicity codes were shown to also achieve list decoding capacity  \cite{kopparty2015list,guruswami-wang}.

Guruswami and Wang \cite{guruswami-wang} later showed that the list of FRS codes (and also of multiplicity codes) is contained in a linear subspace of constant dimension $O(1/\varepsilon)$, which  did not  imply a small list size since the alphabet size of FRS scales linearly with its length. To circumvent this, they  suggested pre-encoding these codes using subspace evasive sets which they also showed to exist. 
Then, Dvir and Lovett were able to  provide an explicit construction of such sets, thus leading to constructions of non-linear subcodes of FRS and multiplicity codes with list size  of $(1/\varepsilon)^{O(1/\varepsilon)}$. 

Lastly, \cite{kopparty-improved-list-size} showed that FRS and multiplicity codes inherently possess the capability of list decoding with constant list sizes, without any modifications such as subspace evasive sets.
In particular \cite[Theorem 3.10]{kopparty-improved-list-size} showed  that FRS codes with minimum distance at least $\delta$ are  $(\delta - \varepsilon,\ell, L)$-list recovery with  a  list size  $L=(\ell/\varepsilon)^{O\left(\frac{1}{\varepsilon}\log(\frac{\ell}{1-\delta})\right)}$. In particular, for list decoding , i.e., for $\ell=1$ the list size is $L=(1/\varepsilon)^{O\left(\frac{1}{\varepsilon}\log(1-\delta)\right)}$. Our work extends these results by eliminating the reliance on the code's minimum distance. Specifically, we get the following theorem for list decoding of FRS codes:

\begin{thm}[Informal, see Corollary \ref{frs-cor}]
For an FRS code $C$ with parameters $\frac{16}{\varepsilon^2} \leq s$, and a minimum distance of at least $\delta$, $C$ is $(\delta - \varepsilon, L)$-list decodable with a list size bounded by 
\[ L \leq  \Big( \frac{1}{\varepsilon}\Big)^{\frac{4}{\varepsilon}}. \]
\end{thm}

By adopting the approach of \cite{kopparty-improved-list-size} which uses the result of \cite{subspace-design} on subspace designs, we  extend the result to list recovery and  prove the following.

\begin{thm}[Informal, see Theorem \ref{frs-list-recovery-best-result}]
For an FRS code $C$ with parameters $\frac{16\ell}{\varepsilon^2} \leq s$, and a minimum distance  $\delta$, $C$ is $(\delta - \varepsilon,\ell, L)$-list recoverable with a list size given by 
\[ L =\Big(\frac{\ell}{\varepsilon}\Big)^{O\big(\frac{1+\log \ell}{\varepsilon}\big)}. \]
\end{thm}
To achieve these results, we establish that the list of codewords is necessarily limited in size when confined to a low-dimensional subspace. This approach, while bearing similarities to the method used in \cite{kopparty-improved-list-size}, incorporates different analysis  that facilitates the attainment of the improved bounds. Applying a similar strategy, we derive the following theorem for multiplicity codes:

\begin{thm}[Informal, see Corollary \ref{mult-cor}]
For a multiplicity code $C$ with parameters $\frac{16}{\varepsilon^2} \leq s$, and a minimum distance of at least $\delta$, $C$ is $(\delta - \varepsilon, L)$-list decodable with a list size bounded by 
\[ L \leq  \left(\frac{1}{\varepsilon}\right)^{O\left(\frac{1}{\varepsilon}\right) }. \]
\end{thm}

\begin{remark}
The work of \cite{kopparty-improved-list-size} introduced an efficient randomized algorithm for outputting all codewords on the list with high probability. Our improvements in the list size bound suggest that a minor modification of their algorithm can improve its running time and success probability. Details of the modified algorithm are presented in the proof of Lemma \ref{main-lemma}.
\end{remark}

\vspace{0.2cm}
\noindent{\bf List Size Bounds for a Fixed Decoding Parameter.} We begin by recalling the following result from \cite{linear-algebraic-approach}. 
\begin{thm}[Theorem 7, \cite{linear-algebraic-approach}]
\label{guruswami-thm}
For the Folded Reed-Solomon  code of rate \(R\) and folding parameter \(s\), the following holds for all decoding parameters \(m \in \mathbb{N}\), \( m \leq s\). Given a received word \(y\), one can find a basis for a subspace of dimension at most \(m-1\) that contains all FRS codewords that differ from \(y\) in at most a fraction \(\frac{m}{m+1} \left(1 - \frac{sR}{s-m+1}\right)\) of the coordinates.
\end{thm}
From Theorem~\ref{guruswami-thm}, it is evident that to achieve list decoding capacity, one must allow both \(m\) and \(s\) to increase. 
Instead,  we consider the list decoding problem for a fixed decoding parameter \(m\), and inquire about the implications for the list size. For \(m=1\), Theorem~\ref{guruswami-thm} reduces to a unique decoding algorithm capable of correcting up to a fraction \((1-R)/2\) of errors. Therefore, the first nontrivial case, which also enables decoding beyond half the minimum distance, is for the decoding parameter \(m \geq 2\). The motivation behind this question is as follows: Assume that the primary goal is not to achieve the maximum possible list decoding radius but rather to ensure that the list size is bounded, say at most 2. What then is the largest list decoding radius one can achieve under these constraints?
 
The technical results we obtain rely on a more refined analysis of the bounds presented earlier, which are then applied for a fixed $m$.
Notably, we obtain the following result for decoding parameter \(m=2\).

\begin{thm}(Informal, see Theorem \ref{optimal-list-2})
\label{optimal-list-2-informal}
FRS code of rate \(R\), code length $n$, and  alphabet size \(n^{O(1/\varepsilon)}\) is \(\left(\frac{2}{3}(1-R)-\varepsilon, 2\right)\)-list decodable for any $\varepsilon>0$.  Moreover, there exists an efficient randomized algorithm that outputs the possibly two codewords in the list. 
\end{thm}
Theorem \ref{optimal-list-2-informal} implies that FRS codes asymptotically achieve the generalized Singleton bound \cite{Shangguan2020combinatorial} for a list of size \(2\) over a polynomial alphabet size. As far as we know, this is the first explicit instance of a code realizing this property. It is noteworthy that \cite{brakensiek2023improved} demonstrated that exactly meeting the generalized Singleton bound \cite{Shangguan2020combinatorial} requires an exponential alphabet size. On the other hand, \cite{guo2023randomly} established the existence of RS codes that asymptotically achieve this bound with a polynomial field size. Subsequently, \cite{alrabiah2023randomly} extended their result by showing that it also holds over a linear field size.

For decoding parameter $m>2$, we get the following result on the list size of FRS code.
\begin{restatable}{thm}{mytheorem}
\label{mytheorem}
The Folded Reed-Solomon code of rate \(R\), decoding parameter $m\geq 3$ and folding parameter \(s\geq m\) large enough, is  
$\left(\frac{m}{m+1} \left(1 - \frac{sR}{s-m+1}\right),L\right)$-list decodable 
with 
$$L\leq \begin{cases}
 (m-1)^{m-1}\left(\frac{m}{m-2}\right)^{m-2} & \eqref{cond2} \text{ holds},\\
 \frac{(m+1)^{m-1}}{(1+mR)(1-R)^{m-2}} & \text{ else,}
\end{cases}$$
where     
\begin{equation}
    \label{cond2}
    \frac{m-1}{m+1} + \frac{(m-1)m}{m+1}\frac{sR}{s-m+1} \leq 1
\end{equation}

\end{restatable}

\vspace{0.2cm}
\noindent{\bf Organization.}
The paper is structured as follows: We begin in Section \ref{preliminaries} with notations and preliminaries. Section \ref{section-2} is dedicated to demonstrating that FRS codes and univariate multiplicity codes achieve list decoding capacity with improved bounds on the list size. In Section \ref{sec-improved-for-FRS}, we provide an enhanced list size bound specifically for list recovering FRS codes. Finally, Section \ref{sec-fixed-decding-parameter} discusses bounds on the list size for fixed decoding parameters.

\section{Preliminaries and Notations}
\label{preliminaries}
We start with the needed   formal definitions and notations. For an integer $n\in \mathbb{N}$, let $[n]=\{1,\ldots,n\}$. Throughout this paper, \(\log\) denotes the logarithm taken in base \(2\). The finite field with \(q\) elements is denoted  by \(\Fq\). For a finite alphabet \(\Sigma\) and  two strings \(x,y\) in \(\Sigma^n\), their relative distance 
\[ dist(x, y) = \frac{|\{i \in [n] : x_i \neq y_i\}|}{n}. \]
is the proportion of coordinates where \(x\) and \(y\) differ. 
For a positive integer \(\ell\), the set \(\binom{\Sigma}{\ell}\) includes all subsets of \(\Sigma\) with size \(\ell\). For a string \(x\) in \(\Sigma^n\) and a set \(S\) in \(\binom{\Sigma}{\ell}^n\) 
\[ dist(x, S) = \frac{|\{i \in [n] : x_i \notin S_i\}|}{n}, \]
is the distance between \(x\) and \(S\), i.e.,   the fraction of coordinates \(i\) in \([n]\) for which \(x_i\) is not in \(S_i\).

\subsection{Basics of Error-correcting Codes}
    Consider an alphabet \(\Sigma\) and a positive integer \(n\) representing the block length. A code is  a subset \(C \subseteq \Sigma^n\), with its elements termed codewords. When \(\mathbb{F}\) is a finite field and \(\Sigma\) a vector space over \(\mathbb{F}\), an \(\mathbb{F}\)-linear code \(C \subseteq \Sigma^n\) is a code that forms an \(\mathbb{F}\)-linear subspace of the vector space \(\Sigma^n\).  \(R := \frac{\log |C|}{\log(|\Sigma|^n)}\)is the rate of the code \(C\), which for \(\mathbb{F}\)-linear codes is \(\frac{\dim_\mathbb{F}(C)}{n \cdot \dim_\mathbb{F}(\Sigma)}\). We say that $\delta$ is the minimum (relative) distance  of \(C\) if it is the minimum relative distance between  any two distinct codewords \(c_1, c_2 \in C\),  hence necessarily  \(dist(c_1, c_2) \geq \delta\). 
    
Given a code \(C \subseteq \Sigma^n\) with minimum distance $\delta$, a  
 parameter \(\rho < \frac{\delta}{2}\), and a string \(w \in \Sigma^n\), decoding from a \(\rho\) fraction of errors corresponds to finding the unique \(c \in C\) (if it exists) that satisfies \(dist(c, w) \leq \rho\). This notion is therefore  called unique decoding. 

List decoding extends the notion of unique decoding to the realm of   error-correction  beyond the \(\frac{\delta}{2}\) fraction of errors by allowing to construct  a short list of nearby codewords. Specifically, for a \(\rho \in (0, 1)\) and an integer \(L\), a code \(C \subseteq \Sigma^n\) is \((\rho, L)\)-list decodable if for any \(w \in \Sigma^n\), at most \(L\) distinct codewords \(c \in C\) satisfy \(dist(c, w) \leq \rho\). List recovery,  a generalization of list decoding, involves receiving a set of potential symbols for each coordinate and outputting codewords consistent with many of these sets. Formally, a code \(C \subseteq \Sigma^n\) is \((\rho, \ell, L)\)-list recoverable if, for any \(S \in \binom{\Sigma}{\ell}^n\), no more than \(L\) distinct codewords \(c \in C\) satisfy \(dist(c, S) \leq \rho\). It is clear that list decoding is a special case of list recovery for \(\ell = 1\).

\subsection{Folded Reed Solomon and Multiplicity Codes}

In this section, we introduce the code families that are the focus of this paper. Folded Reed-Solomon (FRS) codes \cite{GR08} and (univariate) 
multiplicity codes referenced in \cite{rosenbloom1997codes,kopparty2015list, guruswami-wang}.

\vspace{0.2cm}
\textbf{Folded Reed-Solomon codes.} Consider a prime power $q$ and nonnegative integers $s, d, n$ with the condition $n \leq \frac{q - 1}{s}$ and  assume that $\alpha \in \mathbb{F}_q$ is a primitive element. 
The code $\text{FRS}_{q,s}(n, d)$, over the alphabet $\mathbb{F}_q^s$, associates each polynomial $P(X) \in \mathbb{F}_q[X]$ of degree at most $d$ with a codeword in $(\Fq^s)^n$ as follows
\[ P(x) \mapsto  \begin{pmatrix}
P(\alpha^0) & P(\alpha^{s}) & \ldots & P(\alpha^{(n-1)s}) \\
P(\alpha^1) & P(\alpha^{s+1}) & \ldots & P(\alpha^{s-1} a_2) \\
\vdots & \vdots & \ddots & \vdots \\
P(\alpha^{s-1}) & P(\alpha^{2s-1}) & \ldots & P(\alpha^{ns-1} )
\end{pmatrix}. \]
Note that  the special case of $s = 1$ corresponds to RS codes. The basic properties of FRS codes are given in the following claim.

\begin{claim}\cite{GR08}  $\text{FRS}_{q,s}(n, d)$ is an $\mathbb{F}_q$-linear code over the alphabet $\mathbb{F}_q^s$, with block length of $n$,  rate $\frac{d + 1}{sn}$, and  minimum relative distance at least of $1 - \frac{d}{sn}$.
\end{claim}

\vspace{0.2cm}
\textbf{Univariate multiplicity codes.} For a  polynomial $P(x)$ over $\mathbb{F}_q$ and an integer $i \in \mathbb{N}$, the $i$'th (Hasse) derivative $P^{(i)}(X)$ is the coefficient of $Z^i$ in the series expansion
\[ P(X + Z) = \sum_i P^{(i)}(X)Z^i. \]
Let $q$ be  a prime power, $s, d, n$ nonnegative integers  with $n \leq q$, and  distinct elements $a_1, a_2, \ldots, a_n\in \mathbb{F}_q$. The   
 code $\text{MULT}_{q,s}(n, d)$, over the alphabet $\mathbb{F}_q^s$, maps each polynomial $P(X) \in \mathbb{F}_q[X]$ of degree at most $d$ to a codeword as follows
\[ P(x) \mapsto  
\begin{pmatrix}
P(a_1) & P(a_2) & \ldots & P(a_n) \\
P^{(1)}(a_1) & P^{(1)}(a_2) & \ldots & P^{(1)}(a_n) \\
\vdots & \vdots & \ddots & \vdots \\
P^{(s-1)}(a_1) & P^{(s-1)}(a_2) & \ldots & P^{(s-1)}(a_n)
\end{pmatrix}. \]
Similar to before, the case of $s = 1$ boils down to  RS codes. 
The properties of multiplicity codes are given next.

\begin{claim} \cite[Lemma 9]{kopparty2014highrate}. The  $\text{MULT}_{q,s}(n, d)$ is an $\mathbb{F}_q$-linear code over the alphabet $\mathbb{F}_{q^s}$, with block length $n$, rate $\frac{d+1}{sn}$, and relative distance at least $1 - \frac{d}{sn}$.
\end{claim}
\section{Bounds on the List Size of Folded Reed-Solomon and Multiplicity Codes}
\label{section-2}
In this section, we show that Folded Reed-Solomon and multiplicity codes are list decodable up to capacity with a smaller list size bound than previously known. Our proof begins in Section~\ref{output-list-is-small}, where we apply an approach similar to that in \cite{kopparty-improved-list-size}. Here, we establish that the list cannot contain a large number of codewords from a low-dimensional subspace. Subsequently, to derive our main results, in Section~\ref{instantiation-to-FRS}, we integrate this result with the results of \cite{guruswami-wang, kopparty-improved-list-size}  that show that the lists of Folded Reed-Solomon and multiplicity codes are contained in a low-dimensional subspace.

\subsection{Output list has small intersection with low dimensional subspaces}
\label{output-list-is-small}

Our main lemma, presented next, can be seen as an improved version of \cite[Lemma 3.1]{kopparty-improved-list-size}. Furthermore, we believe that this new version offers a more streamlined proof. Broadly, the lemma implies that when list recovering a linear code with a large distance, the output list cannot contain many codewords from a low-dimensional subspace. Specifically, for list decoding, this suggests that codewords that are  clustered within a small Hamming ball, must span a large-dimensional subspace.

\begin{lemma}
\label{main-lemma}
Let $\cC \subseteq (\Fq^s)^n$ be a linear code with relative  minimum distance $\delta>0$ that is $(\delta-\varepsilon,\ell,L)$-list recoverable. Assume further that the output list size is contained in subspace $V\subseteq \cC$ of dimension at $r$, then the output list size  
\begin{equation}
\label{eq:main-lemma}
    L\leq \Big(\frac{\ell}{\varepsilon}\Big)^r.
\end{equation}

Moreover, there is a randomized algorithm that, given a basis for $V$, lists recover $\cC$ with the above parameters in time $\text{poly}(\log q, s, n, L)$.

\end{lemma}

\begin{proof}
Let $S=(S_1,\ldots, S_n)\in \binom{\Fq^s}{\ell}^n$, and assume that the elements of each set $S_i$ are arbitrarily ordered. 
For each codeword $c\in \cC$ in the list, i.e., $dist(x, S) \leq 1-\delta+\varepsilon$, we will construct many unique certificates, in the sense that a certificate can not belong to two different codewords. Then the bound on the list size will follow by bounding the possible number of certificates. 
Before proceeding with the certificate construction, we will need the following simple observation. 

\begin{obser}
\label{obser}
Any projection of the subspace $V$ on more than $(1-\delta)n$ coordinates is injective, i.e., the kernel of the projection is trivial, as otherwise, it would contradict the assumption that the minimum distance of the code is at least $\delta n$. 
\end{obser}

\textbf{ Certificates construction:} Consider  a codeword $c\in \cC, dist(c,S)\leq 1-\delta+\varepsilon$ that is in the list, and let $A\subseteq [n]$ be the agreement set, i.e., $A=\{i\in [n]: c_i\in S_i \}$, where $c_i$ is the $i$th coordinate (column) of $c.$ Then, clearly $|A|\geq (1-\delta+\varepsilon)n.$
 
The  certificates, which are   vectors of length $r$ over the set $[n]\times [\ell]$, will be constructed iteratively. Assume that the first $i-1$ entries of the certificate are $((j_1,m_1),\ldots,(j_{i-1},m_{i-1}))$  and we construct the $i$th entry for 
$1\leq i\leq r$. Let $T_0=\{0\}$ be the zero subspace and $T_{i-1}$ be the subspace obtained by  the projection of $V$ on the $i-1$ coordinates $j_1,\ldots,j_{i-1}$.
Clearly $\dim(T_{i-1})\leq r$. If $\dim(T_{i-1})=r$, let  the $i$th certificate entry be $(j,m)$ for some $j\in A$ such that $c_j$ equals the $m$th element of  $S_j$. Hence, there are at least $(1-\delta+\varepsilon)n$ options for the $i$th entry. 
If \(\dim(T_{i-1}) < r\), we select a coordinate \(j \in A\) that increases the dimension of the projection. More precisely, we have
\begin{equation}
\label{yaron1}
\dim(T_{i-1}) < \dim(V_{\{j_1, \ldots, j_{i-1}, j\}}),
\end{equation}
where \(V_{\{j_1, \ldots, j_{i-1}, j\}}\) denotes the projection of \(V\) onto the coordinates \(j_1, \ldots, j_{i-1},j\). We claim that  by  Observation~\ref{obser}, there are at least \(\varepsilon n\) such coordinates \(j\) in \(A\). Indeed, let $T$ be the projection of $V$ onto the coordinates \(j_1, \ldots, j_{i-1}\) and all the coordinates $j$ that do not satisfy \eqref{yaron1}. Then, by definition $\dim(T_{i-1})=\dim(T)<r$. Now, since the projection is non-injective, by Observation~\ref{obser}, $T$ is a projection on at most $(1-\delta)n$ coordinates, which implies that there are at least $\varepsilon n$ such $j$'s in $A$. We then set the \(i\)th entry of the certificate to be \((j, m)\), for one such $j \in A$ and where \(c_j\) equals the \(m\)th element of \(S_j\).
We conclude that for each codeword in the list, one can construct at least $(\varepsilon n)^r$ certificates. On the other hand, there are at most $(\ell n)^r$ possible certificates, therefore assuming that distinct codewords have distinct certificates this implies that the total number of codewords in the list is at most  $(\ell n)^r/(\varepsilon n)^r=(\ell/\varepsilon)^r$. 

We are left to show that distinct codewords have distinct certificates. Indeed, this will follow  by showing that given a certificate of a codewrd one can recover the codeword from which it was constructed. Let $((j_1,m_1),\ldots, (j_r,m_r))$ be a certificate of the codeword $c\in \cC$. Then, note that by construction,  $\min\{r,\dim(T_{i-1})+1\}\leq \dim(T_i)$ and therefore $\dim(T_r)=r$. Hence, $c$ is the only vector in $V$ that satisfies that 
$c_{j_i}$ equals the $m_i$th element of $S_{j_i}$ for $1\leq i\leq r.$

Next, we describe the randomized algorithm that outputs the list.
The algorithm is given  input  lists $S_1,\ldots,S_n\in \Fq^s$ each of size $\ell$, and a basis  for the  $\Fq$-linear subspace $V\subseteq C$ of dimension $r$ containing
the output list.

\vspace{2mm}
\hrule
\vspace{2mm} 
\textbf{Algorithm: {\bf Prune}($S_1,\ldots,S_n$, $V$, $r$)}
\vspace{2mm} 
\hrule
\vspace{2mm} 

\begin{enumerate}
    \item Initialize $\mathcal{L} = \emptyset$
    \item Pick uniformly at random $i_1, i_2, \ldots, i_r \in [n]$
    \item For each choice of $y_1 \in S_{i_1}, y_2 \in S_{i_2}, \ldots, y_r \in S_{i_r}$:
    \begin{itemize}
        \item If there is exactly one codeword $c \in V$ such that $c_{i_j} = y_j$ for all $j \in [r]$, then $\mathcal{L} \gets \mathcal{L} \cup \{c\}$
    \end{itemize}
    \item Output $\mathcal{L}$
\end{enumerate}

\vspace{2mm} 
\hrule
\vspace{2mm} 

It is evident that if the randomly selected coordinates \(i_1, \ldots, i_r\) correspond to the coordinates of a certificate for a codeword \(c\) in the list, then \(c\) will be included in the output of the \textbf{Prune} algorithm. Consequently, each codeword is outputted with a probability of at least \(\varepsilon^r\) in each iteration of \textbf{Prune}. Indeed,  each codeword has at least $(\varepsilon n)^r$ unique certificates, and there are at most $n^r$ options to select the coordinates of the  certificates. Therefore, by running \textbf{Prune} \(O\left(\varepsilon^{-r} \log \left(\frac{\ell}{\varepsilon}\right)^r\right)\) times and aggregating the output lists, we can assert with high probability, by the union bound, that every codeword in the list will be included in the final output.
\end{proof}
\begin{remark}
The {\bf Prune} algorithm described above is nearly identical to the one presented in \cite{kopparty-improved-list-size}, which is why we have adopted the same name. The sole difference between the two versions is in the second step: the {\bf Prune} algorithm in \cite{kopparty-improved-list-size} selects $t=\Omega\big(\frac{r}{\varepsilon(1-\delta)}\log\big( \frac{r}{\varepsilon(1-\delta)}\big)\big)$ coordinates, whereas our version chooses only $r$ coordinates. Furthermore, in \cite{kopparty-improved-list-size}, the probability of outputting a codeword from the list in a single iteration of {\bf Prune} is lower bounded by $(1-\delta)^{O(t)}$, which is significantly smaller than our probability of $\varepsilon^r$. This difference, combined with our improved bound on the list size, suggests that our version of the {\bf Prune} algorithm requires fewer iterations to output all the codewords in the list with high probability. \end{remark}

\subsection{Improved list-size bounds for FRS and  univariate multiplicity codes}
\label{instantiation-to-FRS}
In this section we apply Lemma \ref{main-lemma} together with known results on the list-recoverability of FRS and multiplicity codes, showing that there is an efficient algorithm that outputs a small dimensional subspace that contains all the codewords of the list. We begin by  citing  the result regarding FRS codes.  

\begin{thm}
\cite[Theorem 7]{guruswami-wang}
\label{FRS-output-subspace}
 Let $q$ be a prime
power, and let $s, d, n$ be nonnegative integers such that $n \leq (q -1)/s$. Let $\delta:= 1-d/(sn)$  be a lower
bound on the relative distance of the folded Reed-Solomon code $\text{FRS}_{q,s}(n, d)$. 
 Let $\varepsilon > 0$ and $\ell\in  \mathbb{N}$ be such  that $16\ell/\varepsilon^2\leq s$. Then $\text{FRS}_{q,s}(n, d) \subseteq (\Fq^s)^n$ is $(\delta-\varepsilon,\ell,L)$-list recoverable, where the output
list is contained in an $\Fq$-linear subspace $V$ of  $\text{FRS}_{q,s}(n, d)$ of dimension at most $r = 4\ell/\varepsilon$. 
Moreover, there is a (deterministic) algorithm that outputs a basis for $V$ in time poly$(\log q, s, d, n)$.
\end{thm}

By combining Lemma \ref{main-lemma} and Theorem \ref{FRS-output-subspace}, we get the following main theorem of this section. 

\begin{thm}
\label{main-thm1-frs}
\textbf{(Constant-size output list for FRS).} Let $q$ be a prime power, and let $s, d, n$ be nonnegative integers such that $n \leq \frac{q - 1}{s}$. Let $\delta := 1 - \frac{d}{sn}$ be a lower bound on the relative distance of the folded Reed-Solomon code $\text{FRS}_{q,s}(n, d)$. Let $\varepsilon > 0$ and $\ell \in \mathbb{N}$ be such that $\frac{16\ell}{\varepsilon^2} \leq s$. Then $\text{FRS}_{q,s}(n, d)$ is $(\delta - \varepsilon, \ell, L)$-list recoverable with
\[ L \leq  \Big( \frac{\ell}{\varepsilon}\Big)^{\frac{4\ell}{\varepsilon}}. \]
Moreover, there is a randomized algorithm that lists recover $\text{FRS}_{q,s}(n, d)$ with the above parameters in time $\text{poly}(\log q, s, d, n, L)$.
  
\end{thm}

By specializing the above theorem for list decoding, i.e., $\ell=1$, we get the following corollary.

\begin{cor}
\label{frs-cor}
 For the above parameters, the $\text{FRS}_{q,s}(n, d)$ is $(\delta - \varepsilon, L)$-list decodable with list size    
 \[ L \leq  \Big( \frac{1}{\varepsilon}\Big)^{\frac{4}{\varepsilon}}. \]
\end{cor}

Next, we turn to apply Lemma \ref{main-lemma} to multiplicity codes. But first, we will need the following theorem, which appeared in \cite{kopparty-improved-list-size}. 

\begin{thm}\cite[Thoreom 3.7]{kopparty-improved-list-size}
\label{Mult-subspace-lemma}
 Let $q$ be a prime power, and let $s, d, n$ be nonnegative integers such that $n \leq q$. Let $\delta := 1 - \frac{d}{sn}$ be a lower bound on the relative distance of the multiplicity code $\text{MULT}_{q,s}(n, d)$. Let $\varepsilon > 0$ and $\ell \in \mathbb{N}$ be such that $\frac{16\ell}{\varepsilon^2} \leq s$ and $\frac{4\ell}{\varepsilon} \leq \text{char}(\mathbb{F}_q)$. Then $\text{MULT}_{q,s}(n, d)$ is $(\delta - \varepsilon, \ell, L)$-list recoverable, where the output list is contained in an $\mathbb{F}_q$-linear subspace $V \subseteq \text{MULT}_{q,s}(n, d)$ of dimension at most $\frac{4\ell}{\varepsilon} \left( 1 + \frac{d}{\text{char}(\mathbb{F}_q)} \right)$.

Moreover, there is a (deterministic) algorithm that outputs a basis for $V$ in time $\text{poly}(\log q, s, d, n)$.

\end{thm}

By combining Lemma \ref{main-lemma} with Theorem \ref{Mult-subspace-lemma} we get the following result. 

\begin{thm}
\label{mult-main-thm-1}
Let $q$ be a prime power, and let $s, d, n$ be nonnegative integers such that $n \leq q$. Let $\delta := 1 - \frac{d}{sn}$ be a lower bound on the relative distance of the univariate multiplicity code $\text{MULT}_{q,s}(n, d)$. Let $\varepsilon > 0$ and $\ell \in \mathbb{N}$ be such that $\frac{16\ell}{\varepsilon^2} \leq s$ and $\frac{4\ell}{\varepsilon} \leq \text{char}(\mathbb{F}_q)$. Then $\text{MULT}_{q,s}(n, d)$ is $(\delta - \varepsilon, \ell, L)$-list recoverable with
\[ L \leq  \left(\frac{\ell}{\varepsilon}\right)^{\frac{4\ell}{\varepsilon} \left( 1 + \frac{d}{\text{char}(\mathbb{F}_q)} \right)}. \]
Moreover, there is a randomized algorithm that lists recover $\text{MULT}_{q,s}(n, d)$ with the above parameters in time $\text{poly}(\log q, s, d, n, L)$.

\end{thm}
By specializing the above theorem for list decoding, i.e., $\ell=1$, we get the following corollary.

\begin{cor}
\label{mult-cor}
    For the above parameters, the $\text{MULT}_{q,s}(n, d)$ is $(\delta - \varepsilon, L)$-list decodable with list size 
\[ L \leq  \left(\frac{1}{\varepsilon}\right)^{\frac{4}{\varepsilon} \left( 1 + \frac{d}{\text{char}(\mathbb{F}_q)} \right)}. \]

\end{cor}

\section{Improved Bound on the List Size of FRS Codes}
\label{sec-improved-for-FRS}
Similarly to \cite{kopparty-improved-list-size}, in this section, we present an improved bound on the list size for the problem of list-recoverability for FRS  codes. The intuition behind this improvement is as follows: The exponent in the bound of Lemma~\ref{main-lemma} is derived from the length $r$ of the certificates. Recall that in the proof of the lemma, each new coordinate of the certificate increased the dimension of the projection by at least one. Therefore, after $r$ steps, we were guaranteed that the dimension of the projection is $r$, i.e., the projection is injective. However, this assumption is conservative, as each new coordinate of the code is, in fact, a vector of length $s$. Consequently, it is plausible to assume that in some cases, the dimension increases by more than one. Thus, fewer than $r$ steps may be required to guarantee that the projection is injective.

In other words, we can produce much shorter certificates, which immediately imply an improved bound on the list size, as needed. To facilitate this approach, we will utilize the following result, originally proved in the context of the construction of subspace designs \cite{subspace-design}.
Note that the theorem was  proved in \cite{subspace-design}, however we present the theorem as it was articulated in \cite{kopparty-improved-list-size}, employing more suitable terminology for our purposes.

\begin{thm}
\label{subspace-design-FRS}
(\cite[Theorem 14]{subspace-design} see also \cite[Theorem 3.12]{kopparty-improved-list-size}) 
    Let $q$ be a prime power, and let $s, d, n$ be nonnegative
integers such that $n\leq (q-1)/s$. Let $\delta:= 1 -d/(sn)$ be a lower bound on the relative distance of the
folded Reed-Solomon code $\text{FRS}_{q,s}(n, d)$. Let $V\subseteq$$\text{FRS}_{q,s}(n, d)$ be an $\Fq$-linear subspace of dimension
$r <  s$. For $i \in  [n]$, let
$$H_i =\{v \in  (\Fq^s)^n : v_i = 0\}.$$
Then
$$\mathbb{E}_{i\in[n]} [\dim(V \cap H_i)]\leq \frac{1-\delta}{1-r/s}\cdot r.$$
\end{thm}
By the above theorem we prove the following lemma, which will be applied iteratively in the proof of our main result of this section. 
\begin{lemma}
\label{iterative-lemma}
    Let $A\subseteq [n]$ be a subset of size at least $(1-\delta+\varepsilon)n$ and $U\subseteq \text{FRS}_{q,s}(n, d)$ be an $\Fq$ linear subspace of dimension $\dim(U)\leq r< s$. If $r/s\leq \varepsilon/4$ then at least $\varepsilon/4$ fraction of the $i$'s in $A$ satisfy   
    $$\dim(U\cap H_i )\leq \lfloor\beta \cdot \dim(U)\rfloor,$$
where     $\beta:=\frac{(1-\delta)(1+\frac{\varepsilon}{2})}{(1-\delta+\varepsilon)(1-r/s)}$.
\end{lemma}

\begin{proof}
    By Theorem \ref{subspace-design-FRS} 
    $$\mathbb{E}_{i\in A}[\dim(U\cap H_i)]\leq \frac{1}{|A|}\sum_{i=1}^n
    \dim(U\cap H_i)\leq \frac{n}{|A|}\frac{1-\delta}{1-r/s}\cdot \dim(U)\leq 
    \frac{1-\delta}{(1-\delta+\varepsilon)(1-r/s)}\cdot \dim(U).
    $$
    Since $r/s\leq \varepsilon/4$ one can verify that  $\beta<1$, 
    and then, by Markov inequality
\begin{align*}
\mathbb{P}_{i\in A}(\dim(U\cap H_i)\geq \beta\cdot \dim(U))
&\leq
\mathbb{P}_{i\in A}(\dim(U\cap H_i)\geq (1+\varepsilon/2)\mathbb{E}_{i\in A}(\dim(U\cap H_i)))
\\&\leq \frac{1}{1+\varepsilon/2}.
\end{align*}
Therefore, in at least $1-\frac{1}{1+\varepsilon/2}\geq \varepsilon/4$ fraction of the indices $i$ in $A$, $\dim(U\cap H_i)\leq \lfloor\beta \cdot \dim(U)\rfloor$, where  
the floor operation follows by integrality. 
\end{proof}
Equipped with the above lemma we are ready to prove the following lemma which can be viewed as an improvement for Lemma \ref{main-lemma}.

\begin{lemma}
\label{main-lemma-2}
Let $\cC \subseteq (\Fq^s)^n$ be a linear code with relative  minimum distance $\delta>0$ that is $(\delta-\varepsilon,\ell,L)$-list recoverable. Assume further that the output list size is contained in a subspace $V\subseteq \text{FRS}_{q,s}(n, d)$ of dimension at $r<s$. Then, there exists an integer $m\leq r$ such that  the output list size  
$$L\leq \Big(\frac{4\ell}{\varepsilon(1-\delta+\varepsilon)}\Big)^m.$$ 
\end{lemma}

Note that similar to Lemma \ref{main-lemma} also Lemma \ref{main-lemma-2} implies an efficient randomized algorithm that given a basis for $V$, lists recover $\cC$. The algorithm is almost identical to Algorithm {\bf Prune} in Section \ref{output-list-is-small}, so we omit the details.

\begin{proof}
Let $S=(S_1,\ldots, S_n)\in \binom{\Fq^s}{\ell}^n$, and assume that the elements of each set $S_i$ are arbitrarily ordered. 
As in the proof of Lemma \ref{main-lemma}, for each codeword in the list we will construct many unique certificates.

\vspace{0.3cm}
\noindent\textbf{ Certificates construction:} 
 Consider  a codeword $c\in \cC, dist(c,S)\leq 1-\delta+\varepsilon$ that is in the list, and let $A\subseteq [n]$ be the agreement set, i.e., $A=\{i\in [n]: c_i\in S_i \}$ where,  by definition $|A|\geq (1-\delta+\varepsilon)n.$
 
The certificates are vectors of length $m$ over the set $[n] \times [\ell]$, where the exact value of $m$ will be determined later. As in the proof of \Cref{main-lemma}, the entries of the certificate are constructed iteratively. Define the subspace $V_0 := V$ and assume that $i-1$ entries of the certificate have already been constructed, along with the subspace $V_{i-1}$ whose dimension is at most $f^{(i-1)}(r)$, for $1 \leq i \leq m$. Here, $f(x) := \lfloor \beta x \rfloor$, and for an integer $n \geq 0$, $f^{(n)}$ denotes the $n$-fold composition of $f$ with itself.

By Lemma \ref{iterative-lemma} with $U = V_{i-1}$ and the set $A$, for at least a fraction $\frac{\varepsilon}{4}$ of the $j$'s in $A$, we have $\dim(V_{i-1} \cap H_j) \leq f(\dim(V_{i-1}))$. Choose an arbitrary such $j$ and define the $i$th certificate entry to be $(j, k)$, where $c_j$ is equal to the $k$th element of $S_j$. Finally, define the subspace $V_i := V_{i-1} \cap H_j$. It is evident from the choice of $j$ that $\dim(V_i) \leq f(\dim(V_{i-1})) \leq f^{(i)}(r)$, where the last inequality follows from the fact that $f$ is an increasing function.

Notice that since $\beta < 1$, we have $\dim(V_i) < \dim(V_{i-1})$. Therefore, after a finite number of steps, $V_i$ becomes the trivial subspace $\{0\}$. We set $m$ to be a sufficiently large integer such that $V_m = \{0\}$ is guaranteed, regardless of the choices made for the coordinates in the certificate. We choose $m$ to ensure that all certificates have the same length. Clearly, such an $m$ exists; one can initially take $m = r$, although we will later demonstrate that an even smaller $m$ is possible.

Since each entry of the certificate has at least $\varepsilon(1 - \delta + \varepsilon)n/4$ options, we can construct at least $\left(\varepsilon(1 - \delta + \varepsilon)n/4\right)^m$ different certificates. Furthermore, we assert that no two codewords share a certificate. Indeed, given a certificate $((j_1, k_1), \ldots, (j_m, k_m))$ of a codeword $c$, one can uniquely recover $c$ from it. First, observe that the projection of the subspace $V$ onto the coordinates $j_1, \ldots, j_m$ is injective. Otherwise, it would contradict the assumption that $V_m = \{0\}$. Therefore, we conclude that $c$ is the unique vector in the subspace $V$ satisfying $c_{j_i}$ equals the $k_i$th element of $S_{j_i}$ for $i = 1, \ldots, m.$

Finally, since there at  most $(\ell n)^m$ possible certificates, and each codeword in the list has at least $(\varepsilon(1-\delta+\varepsilon)n/4)^m$ certificates, the number of codewords in the list is at most 
\begin{equation}
\label{list-bound}
L\leq \Big(\frac{4\ell}{\varepsilon(1-\delta+\varepsilon)}\Big)^m.
\end{equation}
\end{proof}
To show that one can take $m$ to be a small integer we will need the following lemma.

\begin{lemma}
\label{calculated-exponent}
    For two positive integers $x,y$ and an integer $m\geq x +\log_{\beta}\frac{1}{y}$, $f^{(m)}(xy)=0.$
\end{lemma}

\begin{proof}
Notice that for positive integers $a,b$ 
\begin{equation}
    \label{asdf}
    f^{(a+b)}(c)\leq f^{(a)}(c\beta^b).
\end{equation}
 Therefore
    $$0\leq f^{(m)}(xy)\leq f^{\left(x+\left\lceil\log_{\beta}\frac{1}{y}\right\rceil\right)}(xy)\leq
f^{(x)}(xy\beta^{\left\lceil\log_{\beta}\frac{1}{y}\right\rceil})\leq f^{(x)}(xy\beta^{\log_{\beta}\frac{1}{y}})=
    f^{(x)}(x)=0,$$
where the third inequality follows from \eqref{asdf}. 
  The fourth inequality follows since $f$ is monotonically increasing. The last equality follows since $f(0)=0$  and that for a positive integer $x$, $f(x)\leq x-1$.
\end{proof}
By combining Lemma \ref{main-lemma-2}, Lemma \ref{calculated-exponent} and Theorem \ref{FRS-output-subspace} we get the following main theorem of this section. 

\begin{thm}
\label{frs-list-recovery-best-result}
 Let $q$ be a prime power, and let $s, d, n$ be nonnegative integers such that $n \leq \frac{q - 1}{s}$. Let $\delta := 1 - \frac{d}{sn}$ be a lower bound on the relative distance of the folded Reed-Solomon code $\text{FRS}_{q,s}(n, d)$. Let $\varepsilon > 0$ and $\ell \in \mathbb{N}$ be such that $\frac{16\ell}{\varepsilon^2} \leq s$. Then $\text{FRS}_{q,s}(n, d)$ is $(\delta - \varepsilon, \ell, L)$-list recoverable with
\[ L \leq  \Big(\frac{4\ell}{\varepsilon(1-\delta+\varepsilon)}\Big)^{\frac{1}{\varepsilon}+\log_{1/\beta}4\ell}=\Big(\frac{4\ell}{\varepsilon(1-\delta+\varepsilon)}\Big)^{O\big(\frac{1+\log \ell}{\varepsilon}\big)}. \]
  \end{thm}

\begin{proof}
By Theorem \ref{FRS-output-subspace} the dimension $r$  of the output subspace is at most $4\ell/\varepsilon$, therefore by Lemma \ref{calculated-exponent} for an integer   $m\geq \frac{1}{\varepsilon}+\log_{\beta}\frac{1}{4\ell}$,
$f^{(m)}(4\ell/\varepsilon)=0$, and thus  $V_m=\{0\}$ in the proof of Lemma \ref{main-lemma-2}.  Combining this with   \eqref{list-bound} we get that the list size is at most

 $$\Big(\frac{4\ell}{\varepsilon(1-\delta+\varepsilon)}\Big)^{\frac{1}{\varepsilon}+\log_{1/\beta}4\ell}=\Big(\frac{\ell}{\varepsilon}\Big)^{O\big(\frac{1+\log \ell}{\varepsilon}\big)},$$
 where the equality follows since 
 $\log(1/\beta)=\Omega(\varepsilon)$. 
 Indeed,
 $$\ln(1/\beta)\geq \ln\frac{(1-\delta+\varepsilon)(1-\varepsilon/4)}{(1-\delta)(1+\varepsilon/2)}
 =\ln(1+\frac{\varepsilon}{1-\delta})+\ln(1-\frac{3\varepsilon/4}{1+\varepsilon/2})\geq 
 0.9\frac{\varepsilon}{1-\delta}-1.1\frac{3\varepsilon/4}{1+\varepsilon/2}=\Omega(\varepsilon),
 $$
 where the first inequality follows since $r/s\leq \varepsilon/4$ and the second inequality follows for small enough $\varepsilon>0.$ 
 \end{proof}

\begin{remark}
Similar to FRS, a  tighter bound on the list size of univariate multiplicity codes can be obtained in a similar fashion via applying similar ideas and   an  analogous Theorem  from \cite{subspace-design}. We omit the details.
\end{remark}

\section{List Size Bounds for a Fixed Decoding Parameter}
\label{sec-fixed-decding-parameter}
In this section, we turn our attention to the analysis of the list size for a fixed decoding parameter (See Theorem \ref{guruswami-thm} for its definition).
The technical results we obtain rely on a more refined analysis of the bound given in Lemma \ref{main-lemma}, offering, in some sense, an improved bound on the list size. However, this improvement is contingent upon introducing additional assumptions  (see Lemma \ref{main-lemma-finer-analysis} for more details). Subsequently, in Sections \ref{m=2} and \ref{m>2}, we apply this bound to derive bounds on the list size for a fixed decoding parameter $m$. We specifically focus on the two smallest decoding parameters (\(m=2,3\)), for which bounds on the list size were previously unknown.
 
We note that, although our focus in this section is on the problem of list decoding of FRS codes, the results can also be extended to multiplicity codes and to the problem of list recovery. We begin with some needed definitions.

Given a subspace \(V \subseteq (\mathbb{F}_q^s)^n\) of dimension \(r\), we define a vector \(u = (u_1, \ldots, u_t) \in [n]^t\) as valid if, for any \(j = 1, \ldots, t\),
\[
\dim(V_{u^j}) \geq \min\{r, \dim(V_{u^{j-1}}) + 1\},
\]
where \(u^j = (u_1, \ldots, u_j)\) is the prefix of \(u\) of length \(j\), and \(V_{u^j}\) is the projection of \(V\) onto the coordinates \(u^j\). This definition implies that each coordinate \(j\) in the vector \(u\) increases (if possible) the dimension of the projection of \(V\) onto the coordinates in the prefix of \(u\). Recalling the construction of the certificate in the proof of Lemma~\ref{main-lemma}, we observe that each certificate corresponds to a valid vector of length \(r\).

Next, for a valid vector \(u = (u_1, \ldots, u_t)\), let 
\[
\overline{u} = \max\{|I| : u \subseteq I \subseteq [n], \dim(V_I) = \dim(V_u)\},
\]
where, by an abuse of notation, \(u \subseteq I\) implies that each entry of \(u\) is contained in the set \(I\). Intuitively, \(\overline{u}\) represents the number of coordinates of \(V\) (including the coordinates \(u_1, \ldots, u_t\)) whose values are determined once the values of the coordinates \(u_1, \ldots, u_t\) are fixed.
Lastly, we define two sequences of integers for \(0 \leq i < r\):
\[
r_i = \min \{\overline{u} : u \in [n]^i \text{ is valid} \} \quad \text{and} \quad R_i = \max \{\overline{u} : u \in [n]^i \text{ is valid} \}.
\]
We begin with some basic properties of these sets of integers. The proofs of these properties are trivial and are therefore omitted.

\begin{lemma}
\label{basic-properties}
    \begin{enumerate}
        \item \(r_0 = R_0 = |\{i \in [n] : V_i = \{0\}\}|\), where $V_i$ is the projection of $V$ onto the $i$th coordinate. 
        \item The sequences \(r_i\) and \(R_i\) are increasing with respect to \(i\), and satisfy \(i \leq r_i \leq R_i\).
        \item If \(V\) is contained in a code with a minimum distance  \(\delta n\), then \(R_{r-1} \leq n(1 - \delta)\).
    \end{enumerate}
\end{lemma}
Using the above definitions, we can state the following lemma, which can be viewed as an improvement of Lemma \ref{main-lemma}.

\begin{lemma}
\vspace{5pt} 
\label{main-lemma-finer-analysis}
Let \(\mathcal{C} \subseteq (\mathbb{F}_q^s)^n\) be a linear code with relative minimum distance \(\delta > 0\) that is \((1-\frac{e}{n}, \ell, L)\)-list recoverable. Assume further that the output list size is contained in a subspace \(V \subseteq \mathcal{C}\) of dimension at most \(r\). Then, the output list size 
    \begin{equation}
    \label{eq:fine-bound}
        L \leq \ell^r \prod_{i=0}^{r-1} \frac{n-r_i}{e-R_i}
            \leq \ell^r \frac{(n-r_0)^r}{(e-r_0)(e-(1-\delta)n)^{r-1}},
    \end{equation}
    where \(r_i, R_i\) for \(0 \leq i < r\) are the corresponding parameters for the subspace \(V\).
\end{lemma}
\begin{remark}
     In Sections \ref{m=2} and \ref{m>2}, we apply the less tight bound established in \eqref{eq:fine-bound} to obtain list size bounds for FRS codes.
We note that by analysing the specific parameters  $r_i$ and $R_i$ for FRS one might possibly derive even tighter bounds by applying the first inequality in \eqref{eq:fine-bound}. Lastly, the generality of  the bound in \eqref{eq:fine-bound} implies it is applicable to a broad range of codes that share the above properties and not only to FRS codes.

\end{remark}
\begin{proof}
The proof closely follows the approach of Lemma~\ref{main-lemma}, but with refined bounds on the set containing the certificates and the number of certificates constructible for each codeword in the list.

Consider the procedure of certificate construction for the codewords in the list, as outlined in the proof of Lemma~\ref{main-lemma}. We claim the following:

\begin{enumerate}
    \item The set of all certificates lies in a set of size at most \(\ell^r \prod_{i=0}^{r-1} (n-r_i)\). By construction, each vector of coordinates of length \(r\) derived from a certificate is a valid vector. Therefore, the claim follows by showing that the number of valid vectors of length \(r\) is at most \(\prod_{i=0}^{r-1} (n-r_i)\). Assume that \(i\) coordinates of the valid vector have already been determined, and we need to determine its \((i+1)\)th coordinate. By the definition of \(r_i\), once the values of \(i\) coordinates in a vector of \(V\) are fixed, the values of at least \(r_i\) coordinates become known. Consequently, there are at most \(n-r_i\) coordinates whose selection will increase the dimension of the projection, leading to at most \(n-r_i\) options for the \((i+1)\)th entry of the valid vector.
    \item For each codeword in the list, one can construct at least \(\prod_{i=0}^{r-1} (e-R_i)\) certificates. Similar to the previous claim, assume that \(i\) entries of the certificate have already been constructed. Then, by the definition of \(R_i\), once the values of \(i\) coordinates in a vector of \(V\) are fixed, the values of at most \(R_i\) coordinates become known. Since the number of coordinates each codeword agrees with the received word is lower bounded by \(e\), among these at least \(e\) agreement coordinates, there are at least \(e - R_i\) coordinates whose selection will increase the dimension of the projection.
\end{enumerate}
Combining claims (1) and (2) leads to the first inequality of \eqref{eq:fine-bound}. The second inequality follows from the first inequality and by noting that by Lemma \ref{basic-properties} \(r_0 = R_0\), \(r_0 \leq r_i\), and \(R_i \leq (1-\delta)n\) for all \(i\).
\end{proof}

\begin{remark}
    Lemma \ref{main-lemma-finer-analysis} is indeed  an improvement of Lemma \ref{main-lemma},  as \eqref{eq:main-lemma} follows by the first inequality of  \eqref{eq:fine-bound} with 
$e=(1-\delta+\varepsilon)n$, and by setting $r_i=0$ and $R_i=(1-\delta)n$ for each $i$. 
\end{remark}

We proceed to  apply the bound \eqref{eq:fine-bound} on the list size of $\text{FRS}_{q,s}(n, d)$ for decoding parameter $m=2,3$. Recall that $R=\frac{d+1}{sn}$ is the code rate and  $\delta=1-\frac{d}{sn}$ is a lower bound on the minimum distance of the code.

\subsection{List Decoding Parameter \texorpdfstring{$m=2$}{m=2}}

\label{m=2}
By Theorem~\ref{guruswami-thm} for \(m=2\), the output subspace is of dimension at most \(r=1\), and by the second inequality of \eqref{eq:fine-bound}, the number of codewords in the list is at most 
\(\frac{n-r_0}{e-r_0}\), as we are only considering the list decoding problem in this section, i.e., $\ell=1$. Since \(n>e\), the bound on the list size is an increasing function of \(r_0\). Furthermore, by Lemma~\ref{basic-properties} items (1) and (3), we have that \(r_0=R_0\leq (1-\delta)n<Rn\). Hence, the number of codewords is at most \(\frac{n-Rn}{e-Rn}\). Substituting \(e=n-\frac{m}{m+1} \left(1 - \frac{sR}{s-m+1}\right)n\) from \Cref{guruswami-thm} and \(m=2\), one gets that 
\[
L \leq \frac{3(1-R)}{1-R+\frac{2R}{s-1}} < 3
\]
for any \(2  \leq s\). Hence, the list size is at most 2, whereas the decoding radius by Theorem~\ref{guruswami-thm} is \(\frac{2}{3}(1-R-\frac{R}{s-1})\). The following theorem follows immediately by combining the above discussion with Theorem~\ref{guruswami-thm}, Lemma~\ref{main-lemma} and \Cref{main-lemma-finer-analysis}.

\begin{thm}
\label{optimal-list-2}
Let \(\varepsilon > 0\), then the Folded Reed-Solomon code of rate \(R\), length \(n\) and folding parameter \(s = 1 + \left\lceil \frac{2R}{3\varepsilon} \right\rceil\) is a code over an alphabet size \(n^{O(1/\varepsilon)}\) that is \(\left(\frac{2}{3}(1-R)-\varepsilon, 2\right)\)-list decodable. Moreover, there exists an efficient randomized algorithm that outputs the possibly two codewords in the list. 
\end{thm}

\begin{proof}
    By the above discussion, the decoding radius is \(\frac{2}{3}(1-R-\frac{R}{s-1})\), which is at least \(\frac{2}{3}(1-R)-\varepsilon\) by the choice of \(s\). Furthermore, the list size is guaranteed to be at most 2.
\end{proof}
\subsection{List Decoding Parameter \texorpdfstring{$m>2$}{m>2}}
\label{m>2}
Note first that the dimension of the subspace \(r \geq 2\) for \(m > 2\), and that the right-hand side (RHS) of the second inequality of \eqref{eq:fine-bound} is an increasing function in the variable \(r_0\) at the point \(r_0 = 0\) if and only if \(r \cdot e \leq n\). Since $r=m-1$ this is Equivalent to
\begin{equation}
\label{eq:condition}
    \frac{m-1}{m+1} + \frac{(m-1)m}{m+1}\frac{sR}{s-m+1} \leq 1.
\end{equation}
Notice that whether \eqref{eq:condition} holds or not depends on the parameters of the code \(R, s\) and the decoding parameter \(m\). Our analysis proceeds in two cases, depending on whether \eqref{eq:condition} holds or not.

\vspace{10pt} 
\textbf{Case 1 - \eqref{eq:condition} holds:} In this case, one can verify that the maximum of the RHS of the second inequality of \eqref{eq:fine-bound} is attained at \(r_0 = \frac{re-n}{r-1}=\frac{(m-1)e - n}{m-2}\). Substituting this into \eqref{eq:fine-bound} and replacing \(1-\delta\) with the larger quantity \(R\), we obtain the following bound on the list size:

\[
(m-1)^{m-1}\left(\frac{n-e}{(m-2)(e-Rn)}\right)^{m-2} \approx (m-1)^{m-1}\left(\frac{m}{m-2}\right)^{m-2},
\]
where the \(\approx\) sign follows for large \(s\) compared with \(m\).

\vspace{10pt} 
\textbf{Case 2 - \eqref{eq:condition} does not hold:}
In this case, the RHS of the second inequality of \eqref{eq:fine-bound} is a decreasing function for \(r_0 \geq 0\), and therefore the maximum is attained at \(r_0 = 0\). Hence, by replacing $1-\delta$ with the larger quantity $R$ the bound on the list size becomes 
\[
\frac{n^{m-1}}{e(e-Rn)^{m-2}} = \frac{(m+1)^{m-1}}{\left(1+\frac{smR}{s-m+1}\right)\left(1-\frac{s-m^2+1}{s-m+1}R\right)^{m-2}} \leq 
\frac{(m+1)^{m-1}}{(1+mR)(1-R)^{m-2}}.
\]
By the above, Theorem \ref{mytheorem} follows immediately and is restated again for convenience. 
\mytheorem*

\begin{example}
    If \eqref{eq:condition} holds for \(m=3\), the list size is at most \(12\) for a large enough folding parameter \(s\). Otherwise, the list size is at most \(\frac{16}{(1+3R)(1-R)}\), and for a rate, e.g., \(R=\frac{1}{3}\), the list size is at most \(12\). In both cases,  the decoding radius approaches \(\frac{3}{4}(1-R)\) for increasing \(s\). It is an interesting open question whether the list size is in fact at most \(3\), which would imply that the code attains asymptotically the generalized Singleton bound also for a list of size \(3\).

\end{example}

\section*{Acknowledgments}
The author extends deep gratitude to Yaron Shany for his insightful comments on an earlier draft of this paper. His engaging discussions on the subject matter have significantly contributed to refining the results presented in this paper on various levels.

\bibliographystyle{alpha}
\bibliography{biblio}

\end{document}